\documentstyle[twoside,fleqn,espcrc2]{article}

\newcommand{\AmS}{{\protect\the\textfont2
  A\kern-.1667em\lower.5ex\hbox{M}\kern-.125emS}}

\title{Logarithmic Corrections to Scaling in the $XY_2$--Model}

\author{R. Kenna%
        \thanks{Supported by EU Human Capital and Mobility 
                Scheme Project No. CHBI--CT94--1125}
        and 
        A.C. Irving\\
~\\
DAMTP, University of Liverpool L69 3BX, England}
       
\begin{document}

\begin{abstract}
  We study the distribution of partition function zeroes for the $XY$--model
in  two  dimensions.  In particular we find the scaling behaviour of the end
of  the distribution  of zeroes in the complex external magnetic field plane
in the thermodynamic limit (the Yang--Lee edge) and the form for the density 
of  these zeroes.  Assuming  that  finite--size  scaling holds, we show that 
there have to exist logarithmic corrections to the leading scaling behaviour 
of  thermodynamic  quantities in  this model.  These logarithmic corrections 
are  also  manifest  in the  finite--size scaling formulae and  we  identify 
them  numerically.  The  method  presented  here  can  be used to  check the  
compatibility  of scaling  behaviour of odd and even thermodynamic functions  
in  other models too.
\end{abstract}

\maketitle

\section{KOSTERLITZ--THOULESS SCALING}

The partition function for the $d$--dimensional $O(n)$ non--linear 
$\sigma$--model is 
\begin{equation}
  Z_L(\beta,h) 
  = 
  \int_{S_{n-1}}{\prod_{i \in \Lambda}{
    d\sigma_i
    e^{\beta\sum_{\langle i,j \rangle} \sigma_i \sigma_j
       + h \sum_i{\sigma_i}}
  }}
,
\end{equation}
where $L$ denotes the linear extent of the lattice $\Lambda$, $\beta$ is 
the  Boltzmann  factor  and  $h$  is  the reduced  external  field.  The 
$n$--component spin $\sigma_i$ has unit modulus. In the  case of $d=n=2$ 
this is the plane rotator or $XY$--model  and  has a phase transition at 
$(\beta,h)=(\beta_c,0)$ which is caused by the binding  and unbinding of 
vortices.

An approximate renormalization group approach indicates unusual scaling 
behaviour  of the thermodynamic functions as  criticality is approached  
from the disordered phase \cite{KT}. In terms of the reduced temperature
$  t = 1 - {\beta} / \beta_c$, the scaling  behaviour of the correlation 
length, susceptibility and  the  specific  heat is given in \cite{KT} as
\begin{eqnarray}
 \xi_\infty(t) & \sim & e^{at^{-\nu}} 
 \quad , \\
 \chi_\infty(t) & \sim & \xi_\infty^{2-\eta_c} 
 \quad , \label{ktchi}
 \\
 C_\infty(t) & \sim & \xi_\infty^{\tilde{\alpha}} + {\rm{constant}}
 \quad ,
 \label{ktcv}
\end{eqnarray}
where for $t > 0$, $\nu = 1/2$, $\eta_c = 1/4$ and $\tilde{\alpha} =-d =-2$.
The aim of this work is to argue that the latter two  scaling  formulae  are
incompatible as they stand. To this end, a method is presented by which  odd 
and even  thermodynamic  functions (like (\ref{ktchi}) and (\ref{ktcv})) can 
be  related  and  expressed  in  terms  of  partition function zeroes. Using 
certain reasonable assumptions regarding finite--size scaling,  it  is shown 
that  there  have  to  exist   multiplicative   logarithmic   corrections to 
(\ref{ktchi}) and (\ref{ktcv}). This  method  can be applied to any model.

\section{LEE--YANG ZEROES}

For  the   $XY_2$--model  the zeroes in the complex external field strength 
plane are on the imaginary axis \cite{LY,DuNe75}.  Expressing the partition 
function in terms of its Lee--Yang zeroes (denoted $z_j(\beta)$), 
\begin{equation}
 Z_L(\beta,h)
 =
 \rho_L(\beta,h) \prod_j(h-iz_j(\beta))
 \quad,
\label{pfz}
\end{equation}
where $\rho_L$ is a non-vanishing function of $h$ related to the   spectral  
density and contributes only to the regular  part  of  the free energy. The 
singular part of the free energy corresponds to
\begin{equation}
 f_L(\beta,h) 
 = 
 L^{-d}
 \sum_j{ \ln{ (h-iz_j(\beta)) } }
\quad .
\label{free}
\end{equation}
This can be written as
\begin{equation}
 f_L(\beta,h)
 =
 \int_{z=-R}^{R}{   \ln{(h-iz)} g_L(\beta,z) } dz
\quad ,
\end{equation}
in which $R$ is an appropriate cutoff and the density of zeroes,
$g_L(\beta,z)$ is given by
\begin{equation}
 g_L(\beta,z)
 =
 L^{-d}
 \sum_j{\delta(z-z_j(\beta))}
 =
 \frac{\partial G_L(\beta,z)}{\partial z}.
\label{cum}
\end{equation}
Here $G_L$ is the cumulative density of zeroes. The distribution of zeroes 
is symmetric in the real $h$--axis and the (cumulative) density  of zeroes 
is zero up to the Yang--Lee edge. This allows one to write  the   singular 
part of the free energy in the thermodynamic limit as
\cite{Abe,KeLa94,letter}
\begin{equation}
 f_\infty(\beta,h) 
 =
 -2 
 \int_{z_1(\beta)}^{R}{\frac{z}{h^2+z^2}}G_\infty(\beta,z)dz
\quad,
\end{equation}
where $z_1(\beta)$ is the position of the Yang--Lee edge.

This leads to an 
expression for $\chi_\infty$ in terms of the cumulative density of zeroes 
and the edge. Following  \cite{Abe,KeLa94} this gives (independent of the 
model under consideration provided it obeys the Lee--Yang theorem)
\begin{equation}
 G_\infty(\beta,z)
 =
 \chi_\infty (\beta,z)
 z_1^2(\beta)
 \Phi\left( \frac{z}{z_1(\beta)}  \right)
\quad ,
 \label{G}
\end{equation}
where   $\Phi$  is  unknown.  The  specific  heat
can also be written in terms of 
$z_1$ and $G_\infty$ \cite{Abe,KeLa94,letter};
\begin{equation}
 C_\infty(\beta)
 =
 -2
 \int_{z_1(\beta)}^R{
  z^{-1}
  \frac{\partial^2 G_\infty}{\partial \beta^2}
  dz
 }
\label{C}
\quad .
\end{equation}

\section{SCALING AND CORRECTIONS}

Assume the following modified Kosterlitz--Thouless (KT) scaling behaviour
for the singular parts of the thermodynamic functions:
\begin{eqnarray}
 \chi_\infty(t) & \sim & \xi_\infty^{2-\eta_c} t^r
 \quad , \label{ktchimod}
 \\
 C_\infty(t) & \sim & \xi_\infty^{\tilde{\alpha}} t^q
 \quad .
 \label{ktcvmod}
\end{eqnarray}
Assume, furthermore, the edge scales as
\begin{equation}
 z_1(t)  \sim  \xi_\infty^{\lambda} t^p
\quad .
\label{tdedge}
\end{equation}
Putting these in (\ref{G}) and (\ref{C}) gives \cite{letter}
\begin{equation}
 \lambda = \frac{1}{2}(\tilde{\alpha}-2+\eta_c),
\quad 
 p  =  \frac{1}{2}(q-r) + 1 + \nu,
\end{equation}
Thus  the  scaling behaviour of the Yang--Lee edge has multiplicative 
logarithmic corrections even if $\chi_\infty$ and $C_\infty$ have not.

\section{FINITE--SIZE SCALING}

Finite--size scaling (FSS) \cite{Fi72} allows    one to find 
the volume dependency of thermodynamic quantities at $\beta_c$
from their thermodynamic limit scaling behaviour.  Its general  form 
(valid in all dimensions including the upper critical one) 
is \cite{KeLa93}
\begin{equation}
 \frac{P_L(0)}{P_\infty(t)}
 =
 F_P
 \left(
       \frac{\xi_L(0)}{\xi_\infty(t)}
 \right)
\quad ,
\label{FSS}
\end{equation}
where $P_L(t)$ is some thermodynamic  function at reduced 
temperature $t$ for a system of extent $L$ and $F_P$ is 
unknown. For the $XY_2$--model    $\xi_L(0)  \propto L$   \cite{Luck}. 
Applying (\ref{FSS}) to  (\ref{ktchimod}) and (\ref{tdedge}) gives
\begin{eqnarray}
 \chi_L(0) 
 & \sim &
 L^{2-\eta_c}
 (\ln{L})^{-\frac{r}{\nu}}
\quad , \\
\label{fsschi}
 z_1(L)
 & \sim &
 L^{\frac{1}{2}\left({\tilde{\alpha}-2+\eta_c}\right)}
 (\ln{L})^{-\frac{1}{\nu}\left(\frac{q-r}{2}+1+\nu\right)}
\quad .
\label{fssedge}
\end{eqnarray}
The finite--volume counterpart of (\ref{G}), relating the susceptibility
and the zeroes, is (from (\ref{free}))
\begin{equation}
 \chi_L(0)
 =
 L^{-d}
 \sum_j{z_j(L)^{-2}}
 \approx
 L^{-d}
 {z_1(L)}^{-2}
\quad ,
\label{chizero}
\end{equation}
where it has been assumed that the lowest 
lying zero has the  dominant effect. This gives
\begin{equation}
 \tilde{\alpha} = -d = -2, \quad 
 q = -2(1+\nu) = -3.
\label{q}
\end{equation}
Thus the scaling behaviour of the singular part of the specific heat 
indeed exhibits multiplicative logarithmic corrections.

\section{NUMERICAL RESULTS}

The above arguments have yielded no information on the odd correction 
exponent $r$. Careful renormalization group  (RG) considerations give 
$r=-1/16$ \cite{sixteenth}. Our task now is to identify $r$ numerically. 
To this end we use the FSS of the Lee--Yang zeroes.  Accepting the KT 
predictions for $\nu$, $\eta_c$ and $\tilde{\alpha}$ for $t>0$, 
(\ref{fssedge}) gives
\begin{equation}
 z_1(L) \sim L^{-\frac{15}{8}}(\ln{L})^r
\quad .
\label{goal}  
\end{equation} 
A Wolff algorithm \cite{Wolff} was used to simulate the $XY$--model 
on square lattices of sizes $L=32,64,128$  and  $256$. The critical
$\beta$--value was found to be $1.11(1)$  from  phenomonological RG 
methods \cite{RoWy}. The results for  the  lowest  lying  zeroes at 
$\beta = 1.11$  are     $z_1(L) = 0.0023348(7)$,    $0.0006350(2)$,
$0.00017279(5)$ and $0.000047062(13)$  for $L=32,64,128$ and $256$. 
Details on the numerics and the  determination of $\beta_c$ and the 
zeroes   will  be given in a forthcoming publication \cite{letter}.
\begin{figure}[htb]
\vspace{5cm}
\includegraphics{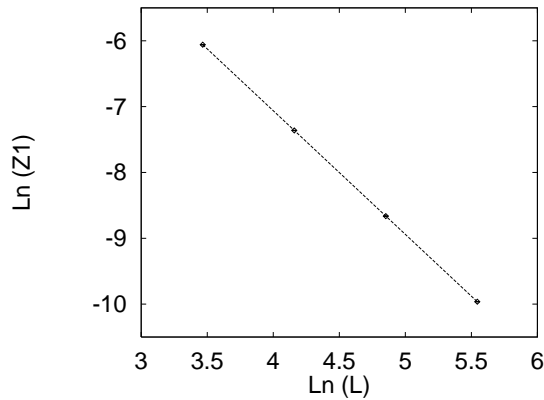}
\caption{Leading FSS of Lee--Yang Zeroes.}
\label{fig:leading}
\end{figure}
\begin{figure}[htb]
\vspace{4.5cm}
\includegraphics{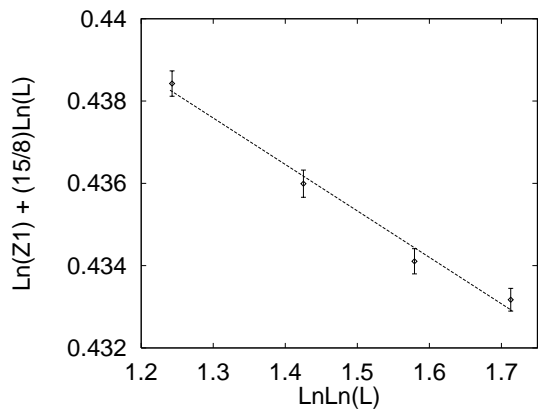}
\caption{Corrections to FSS of Lee--Yang Zeroes.}
\label{fig:corrections}
\end{figure}
In the absence of any corrections, the slope of Figure~\ref{fig:leading} 
should give the leading power--law FSS exponent.    In fact the slope is 
-1.8777(4),  the deviation from the KT value of $-15/8 = -1.875$   being 
due to the presence of logarithmic corrections. To identify  these,  and 
the correction exponent $r$ in  (\ref{goal}),  $\ln{(z_1 L^{15/8})}$  is 
plotted   against  $\ln{\ln{L}}$   in    Figure~\ref{fig:corrections}. A 
straight line is identified.   Its  slope  is $-0.012(1)$.
Thus we have strong evidence for a non--zero value of 
$r$,   albeit    not   in   agreement  with the RG predictions of 
$-1/16 = -0.0625$ from \cite{sixteenth}.

\section{CONCLUSIONS}
 
Theoretical arguments checking the consistency of the scaling
behaviour of odd and even  thermodynamic  functions  at  a KT
phase transition have  been  presented.  The  generally  used
scaling  formulae  have  to  be  modified  by  multiplicative
corrections.   These  are  identified  analytically  for  the 
specific heat and numerically for the   susceptibility.  This 
numerical identification comes via an analysis  of  Lee--Yang 
zeroes,  the   FSS  of  which  is   linked  to  that  of  the 
susceptibility.   \\

We would like to thank P. Lacock for assistance with multihistogram 
reweighting techniques.

\end{document}